\documentclass[twocolumn,showpacs,amsmath,amssymb,nofootinbib,superscriptaddress]{revtex4}

\usepackage{graphicx}
\usepackage{epsfig}
\usepackage{bm}
\usepackage{amsfonts}
\usepackage{amssymb}
\usepackage{color}
\usepackage{float}
\usepackage{amsmath}
\usepackage{dcolumn}
\usepackage{hyperref}
\usepackage{array}
\usepackage{lscape}

\usepackage{soul}
\usepackage{stmaryrd}
\usepackage{graphicx,subfigure,bm,color,psfrag,hyperref}
\usepackage{mathtools}
\usepackage{amsmath,amsthm}
\usepackage{verbatim}
\usepackage[dvipsnames]{xcolor}

\newcommand{\udt}[3]{#1^{#2}_{\phantom{#2}#3}}

\newcommand{\dut}[3]{#1_{#2}^{\phantom{#2}#3}}
\newcommand{\dudt}[4]{#1_{#2\phantom{#3}#4}^{\phantom{#2}#3}}
\def\nabbol{{\mathcal{D}}{}}
\def\Rbol{{\stackrel{\circ}{R}}{}}
\def\Gammabol{{\mathcal{\{ \}}}{}}

\begin{document}
\title{Conformal Gravity and Transformations in the Symmetric Teleparallel Framework}

\author{Viktor Gakis}
\email{vgakis@central.ntua.gr}
\affiliation{Institute of Space Sciences and Astronomy, University of Malta, Msida, Malta}
\affiliation{Department of Physics, National Technical University of Athens, Zografou Campus GR 157 73, Athens, Greece}

\author{Martin Kr\v{s}\v{s}\'ak}
\email{martin.krssak@ut.ee}
\affiliation{Center for Gravitation and Cosmology, College of Physical Sciences and Technology, Yangzhou University, Yangzhou 225009, China}
\affiliation{Laboratory of Theoretucal Physics, Institute of Physics, University of Tartu, Tartu, 50411, Estonia}

\author{Jackson Levi Said}
\email{jackson.said@um.edu.mt}
\affiliation{Institute of Space Sciences and Astronomy, University of Malta, Msida, Malta}
\affiliation{Department of Physics, University of Malta, Msida, Malta}

\author{Emmanuel N. Saridakis}
\email{msaridak@phys.uoa.gr}
\affiliation{Department of Physics, National Technical University of Athens, Zografou Campus GR 157 73, Athens, Greece}
\affiliation{Department of Astronomy, School of Physical Sciences, University of Science and Technology of China, Hefei 230026, P.R. China}

\begin{abstract}

Conformal symmetries appear in many parts of physics and play a unique role in exploring the Universe. In this work we consider the possibility of constructing conformal theories of gravity in the Symmetric Teleparallel Gravity framework, where gravitation is expressed through non-metricity rather than curvature or torsion. We demonstrate that it is possible to formulate a family of conformal theories with second-order field equations and have the metric tensor as the fundamental variable. Moreover, we consider the   addition of a scalar field to the theory, and compare the results to the Teleparallel Gravity setting. Finally, we present the scalar-tensor representation of modified symmetric teleparallel theories, and amongst others we show that there is   an interesting sub-case that has the possibility 
of playing the role of dark energy.
\end{abstract}
\maketitle

\section{Introduction }
The conformal symmetry is usually considered to be only an approximate symmetry of Nature, but recently arguments for considering gravitational theories with full conformal invariance were put forward \cite{Mannheim:2007ki,Hooft:2010ac,Mannheim:2011ds,Hooft:2014daa}. This interest follows primarily from the observation that conformally
invariant theories are expected to be renormalizable and hence a plausible model of quantum gravity, as well as being a plausible method to approach the dark energy and the dark matter problems \cite{Mannheim:2005bfa}. Within the standard framework of Riemannian geometry, used in Einstein's general relativity (GR), the most natural Conformal Gravity model is \textit{Weyl gravity}, where the Lagrangian is considered to be squared in the Weyl tensor \cite{Mannheim:2011ds}. The obvious drawback of this construction is the presence of higher derivative terms in the Lagrangian, leading to a fourth order theory and hence introducing Gauss-Ostrogradsky ghosts. While it has been shown that the ghosts might dissapear in certain situations, they are expected to be present in the full theory \cite{Mannheim:2006rd,Maldacena:2011mk,Anastasiou:2016jix}.

An alternative approach to modified gravity theories, that we follow in the present work, is to consider the torsional formulation of gravity, where instead of the standard Riemannian geometry we use the \textit{teleparallel geometry} \cite{Cai:2015emx}. The connection characterizing the rule of parallel transport of teleparallel geometry is defined by the \textit{teleparallel condition} of vanishing curvature. There are then two distinct cases to be considered.

In the first case that we call here the \textit{torsional teleparallel geometry} the connection is fully determined by an additional condition of the metric compatibility. The non-trivial geometry is then described in terms of the spacetime torsion and allows us to formulate the fully torsion-based theories of gravity. One particular example is the so-called \textit{teleparallel equivalent of general relativity}, which is a theory dynamically equivalent to GR \cite{aldrovandi2012teleparallel,Krssak:2018ywd} originating
from Einstein's attempt for a unified theory \cite{Einstein1928,Sauer:2004hj}, that was also fully developed as an alternative description of GR as early as the 1960s \cite{Moller1961,Hayashi:1967se,Cho:1975dh,Hayashi:1977jd}. The topic was then further explored mainly within the framework of metric-affine theories of gravity \cite{Hehl:1994ue,Obukhov:2002hy}. Recently, various extensions/modifications of teleparallel equivalent of GR, such as $f(T)$ gravity \cite{Ferraro:2006jd,Ferraro:2008ey,Bengochea:2008gz,Linder:2010py,Chen:2010va} and others \cite{Hayashi:1979qx,Kofinas:2014owa,Kofinas:2014daa,Bahamonde:2016kba,Bahamonde:2015zma,Harko:2014aja,Bahamonde:2017wwk,Itin:2016nxk,Hohmann:2017duq,Chen:2019ftv}, became a popular approach to address the accelerated expansion of the Universe, as well as to study gravitational waves \cite{Bamba:2013ooa,Cai:2018rzd,Farrugia:2018gyz,Hohmann:2018jso,Nunes:2018evm}.

The second case that will be of our main interest here are the so-called \textit{symmetric teleparallel gravity} (STG) theories based on the \textit{symmetric teleparallel geometry}, where the connection is defined by the teleparallel condition (vanishing curvature) and the connection being symmetric (zero torsion). As a consequence, the non-trivial geometry is described in terms of the spacetime non-metricity. This geometry was first introduced by Nester and Yo \cite{Nester:1998mp} to formulate the \textit{symmetric teleparallel equivalent of general relativity} (STEGR) and was further developed and extended to various modified scenarios in \cite{Adak:2004uh,Adak:2005cd,Adak:2006rx,BeltranJimenez:2017tkd,BeltranJimenez:2018vdo,Jarv:2018bgs}.

An intriguing feature of teleparallel geometries is that both the torsion and the non-metricity tensors contain only first derivatives
of the dynamical variable, i.e. the tetrad or the metric tensor, which allows us to naturally formulate various extended/modified theories of gravity without introducing higher derivatives of the dynamical variable and hence avoid the problem of Gauss-Ostrogradsky ghosts. In particular, this can be used to formulate a Conformal Gravity theory with second-order field equations, as it was demonstrated by Maluf and Faria \cite{Maluf:2011kf}, who formulated a torsion-based conformally invariant theory of gravity using the standard teleparallel geometry. 

In this work we follow the symmetric teleparallel geometry approach and formulate a non-metricity-based theory, namely the \textit{conformal symmetric teleparallel gravity}. Our work is also related to the recent general work in Ref.~\cite{Iosifidis:2018zwo}, where we focus on the non-metricity approach only and applications to various modified theories of gravity. Similar to the analysis of \cite{Maluf:2011kf}, we investigate two possibilities to implement the conformal symmetry. The first one is the purely geometric Conformal Gravity, where we find that the conformal invariance requires the Lagrangian to be a polynomial quartic in non-metricity tensor, determining its form. Additionally, we investigate the second approach and formulate the conformally invariant theory where the so-called non-metricity scalar, i.e. a term quadratic in non-metricity tensor, is coupled with a scalar field. We find that, unlike the Maluf and Faria case \cite{Maluf:2011kf}, our theory does not posses a physically interesting non-conformal limit. 

The motivation for using symmetric teleparallel geometry to formulate conformal gravity theories stems from the fact that conformal transformations naturally act directly on the metric tensor, which is the fundamental variable of symmetric teleparallel geometry. We argue that this is a more natural setting than the torsional teleparallel geometry approach where the fundamental variable is the tetrad. In this setting, the conformal transformation acts on the tetrad, and hence this symmetry is only an inherited property from the metric tensor.

The outline of the paper is as follows. In Sec.~\ref{sec_intro} we introduce the symmetric teleparallel geometry and the symmetric teleparallel formulation of general relativity, and we show its relation to standard Einstein theory. We then derive the field equations for the general modified STG model. In Sec.~\ref{sec_ngr}, we introduce the \textit{newer general relativity} where the Lagrangian is given by the most general expression quadratic in non-metricity tensor, and we discuss how the conformal transformations act in this theory. In Sec.~\ref{sec_geoconf} we present the \textit{conformal symmetric teleparallel gravity}, extracting the field equations and discussing the conformal invariance introducing a scalar field. In Sec.~\ref{sec_fqb} we investigate the scalar-tensor form of $f(\mathcal{Q})$ and $f(\mathcal{Q},B)$ gravity, as well as of their coupling with a scalar field, and we analyze their conformal transformations. Finally in Sec.~\ref{sec_conclusion} we provide a summary and conclusions.

\textit{Notation}: Throughout the paper we follow the mostly positive convention for the metric. The Levi-Civita connection of the Riemannian geometry is denoted $\Gammabol^\rho{}_{\mu\nu}$ and $\Rbol$ denotes the Ricci curvature scalar of this connection. Moreover, we use the following symbols to distinguish the different covariant derivatives appearing in the paper: $\nabbol_{\mu}$ denotes the covariant derivative with respect to the Levi-Civita connection, $\nabla_\mu$ with respect to the symmetric teleparallel connection, and $\text{D}_\mu$ is the conformal covariant derivative. The symbol $\mathcal{Q}$ is reserved for the STEGR non-metricity scalar, while $Q$ is the general quadratic non-metricity scalar of new general relativity.

\section{Symmetric Teleparallel Gravity and Extensions \label{sec_intro}}

The standard formulation of GR assumes the underlying geometry to be the Riemannian geometry characterized by the symmetric and metric compatible Levi Civita connection with a non-vanishing curvature \cite{Nakahara:2003nw}. However, in principle we are not necessarily limited to the choice of the Levi-Civita connection and we can consider a general linear affine connection of the form \cite{ortin2007gravity,Hehl:1994ue}
\begin{equation}
\Gamma^{\alpha}{}_{\mu\nu}=\Gammabol^{\alpha}{}_{\mu\nu}+\udt{K}{\alpha}{\mu\nu}+\udt{L}{
\alpha}{\mu\nu}\,,\label{Ricci}
\end{equation}
where $\Gammabol^{\alpha}{}_{\mu\nu}$ represents the Christoffel symbols of the Levi-Civita connection, $\udt{K}{\alpha}{\mu\nu}$ is the contortion tension and 
\begin{equation}
\udt{L}{\alpha}{\mu\nu}=\frac{1}{2}g^{\alpha\beta}\left(-Q_{\mu\beta\nu}-Q_{\nu\beta\mu}
+Q_{\beta\mu\nu}\right)\,,\label{disf}
\end{equation}
is the \textit{disformation} tensor which encodes a contribution of the the non-metricity tensor \cite{Nester:1998mp,BeltranJimenez:2017tkd}
\begin{equation}
Q_{\alpha\mu\nu}:=\nabla_{\alpha}g_{\mu\nu}\,,\label{nonmetricity_scalar}
\end{equation}
to the connection. Note that its inverse is given by $Q^{\alpha\mu\nu}=-g^{\alpha\beta}\nabla_{\beta}g^{\mu\nu}$.

The class of teleparallel geometries is characterized by the teleparallel condition of vanishing curvature 
\begin{equation}
R^{\rho}{}_{\sigma\mu\nu}=0\,.
\end{equation}
The standard teleparallel geometry is fully given by the additional condition of metric-compatibility, and can be used to formulate the teleparallel equivalent of GR and its various extensions \cite{Krssak:2018ywd}.

Here we consider the symmetric teleparallel geometry where the connection is given fully by the vanishing curvature (teleparallel condition) and the vanishing torsion (symmetric connection). The most general connection of this kind is the symmetric teleparallel connection 
\begin{equation}
\Gamma^{\alpha}{}_{\mu\nu}=\frac{\partial 
x^{\alpha}}{\partial\xi^{\sigma}}\frac{\partial^{2}\xi^{\sigma}}{\partial x^{\mu}\partial 
x^{\nu}}\,,\label{conn}
\end{equation}
where $\xi^{\sigma}=\xi^{\sigma}(x)$ is an arbitrary function. We can recognize that the connection in \eqref{conn} can be obtained from the zero connection using a coordinate transformation 
\begin{equation}
x^{\mu}\rightarrow\xi^{\mu}(x^{\mu}).
\end{equation}
Therefore, the connection in Eq.~\eqref{conn} turns out to be a pure-gauge connection and hence it is always possible to choose a coordinate system that the connection vanishes. This choice was named as the \textit{coincident gauge} \cite{BeltranJimenez:2017tkd}.

In this way, one can choose to consider a gravitational theory based only on the disformation rather than on the Levi-Civita connection, where the effect of gravity is measured through the non-metricity tensor. In particular one can formulate the \textit{symmetric teleparallel equivalent of general relativity} (STEGR), which is dynamically equivalent to GR for all phenomenology emanating from the field equations, while sustaining differences at the level of the Lagrangian. Let us consider the Lagrangian of the STEGR as 
\begin{equation}
\mathcal{L}_{\text{STEGR}}=\frac{\sqrt{-g}}{16\pi G}\mathcal{Q}\,,\label{lag_stg}
\end{equation}
where 
\begin{equation}
\mathcal{Q}=g^{\mu\nu}\left(\udt{L}{\alpha}{\beta\mu}\udt{L}{\beta}{\nu\alpha}-\udt{L}{
\alpha}{\beta\alpha}\udt{L}{\beta}{\mu\nu}\right)\,,
\end{equation}
is the \textit{non-metricity scalar}.

In order to see that this Lagrangian indeed corresponds to a theory dynamically equivalent to GR, let us write the Einstein-Hilbert Lagrangian as
\cite{Harko:2018gxr} 
\begin{equation}
\mathcal{L}_{EH}=\frac{\sqrt{-g}}{16\pi G}\Rbol=\mathcal{L}_{E} + B\,,\label{EH}
\end{equation}
where $\mathcal{L}_{E}$ represents the Einstein Lagrangian constructed from the Levi-Civita connection \cite{Einstein:1916cd}
\begin{equation}
\mathcal{L}_{E}:=\frac{\sqrt{-g}}{16\pi G}\,g^{\mu\nu}\left(\Gammabol^{\alpha}{}_{\beta\mu}
\Gammabol^{\beta}{}_{\nu\alpha}-\Gammabol^{\alpha}{}_{\beta\alpha}\Gammabol^{\beta}{}_{
\mu\nu}\right)\,,\label{ELag}
\end{equation}
while the total derivative (or boundary term) is given by 
\begin{equation}
B = 
\frac{\sqrt{-g}}{16\pi G}\,\left(g^{\alpha\mu}\nabbol_{\alpha}\Gammabol^{\nu}{}_{\mu\nu}-g^{
\mu\nu}\nabbol_{\alpha}\Gammabol^{\alpha}{}_{\mu\nu}\right)\,,\label{boundary_ein_def}
\end{equation}
where $\nabbol_{\alpha}$ represents the covariant derivative with respect to the Levi-Civita connection. The Einstein-Hilbert Lagrangian \eqref{EH} is the higher derivative version of the equivalent Lagrangian $\mathcal{L}_{E}$ and is ubiquitously adopted due to its covariance, while $\mathcal{L}_{E}$ alone is not covariant within the Levi-Civita connection setting.

Assuming the coincident gauge, namely the case where the connection vanishes, i.e. $\Gamma^{\alpha}{}_{\mu\nu}=0$, the covariant derivative reduces to an ordinary partial derivative, $\nabla_{\alpha}\rightarrow\partial_{\alpha}$, and hence from \eqref{Ricci} it follows that the disformation tensor is essentially the negative of the Christoffel symbols:
\begin{equation}
\udt{L}{\alpha}{\mu\nu}=-\Gammabol^{\alpha}{}_{\mu\nu}\,.
\end{equation}
In this case the STEGR Lagrangian in \eqref{lag_stg} will turn out to coincide with the Einstein Lagrangian \cite{Nester:1998mp}, namely
\begin{equation}
\mathcal{L}_{STEGR}=\mathcal{L}_{E}\,.\label{STEGR_Lagran}
\end{equation}
Therefore, the STEGR Lagrangian produces the exact same field equations as the Einstein-Hilbert Lagrangian, while preserving a different theoretical basis through the non-metricity formalism \cite{Conroy:2017yln}.

Moreover, when we consider an arbitrary coordinate transformation, and hence move away from the coincident gauge since the coordinate change will induce the non-trivial symmetric teleparallel connection in \eqref{conn}, we find that the STEGR Lagrangian \eqref{STEGR_Lagran} -- unlike
the Einstein Lagrangian \eqref{ELag} -- is invariant under diffeomorphisms. Thus, the STEGR can be understood as a covariantization of the Einstein Lagrangian, similarly to the fact that the Einstein-Hilbert Lagrangian can be understood as a covariantization of the same Einstein Lagrangian.
The covariantization here is achieved by considering the non-trivial symmetric teleparallel connection \eqref{conn}, while in the Einstein-Hilbert case this is achieved by adding the higher order boundary $B$-term.

In the case of curvature gravity it is known that modifications can be obtained using various extended forms (such as $f(\Rbol)$ gravity) \cite{Capozziello:2011et}. Hence, in the present framework we can similarly consider the general action \cite{Koivisto:2018aip}
\begin{align}
S_{G} &= \int 
d^{4}x\,\Big[\frac{\sigma}{2}\sqrt{-g}\,f+\dut{\lambda}{\alpha}{\beta\mu\nu}\udt{R}{\alpha}{\beta\mu\nu}+\dut{\lambda}{\alpha}{\mu\nu}\udt{T}{\alpha}{\mu\nu} \nonumber\\
& + \sqrt{-g}\mathcal{L}_{m}\Big]\,,\label{Gen_func_Lagran}
\end{align}
with $\sigma$ the coupling constant for gravitation, and where the Lagrangian assumes a Palatini approach in which $f=f(g_{\mu\nu},\udt{\Gamma}{\alpha}{\mu\nu})$, and the Lagrange multipliers $\dut{\lambda}{\alpha}{\beta\mu\nu}$ and $\dut{\lambda}{\alpha}{\mu\nu}$ are used to exclude the curvature (Riemann) tensor and the torsion tensor from the resulting field equations. A conjugate to $f(g_{\mu\nu},\udt{\Gamma}{\alpha}{\mu\nu})$ can then straightforwardly be defined as \cite{BeltranJimenez:2017tkd}
\begin{equation}
\sqrt{-g}\udt{\mathbb{P}}{\alpha}{\mu\nu}:=\frac{1}{2}\frac{\partial 
(\sqrt{-g}f)}{\partial\dut{Q}{\alpha}{\mu\nu}}\,,\label{conj_Lag_ten}
\end{equation}
and then (\ref{Gen_func_Lagran}) yields the metric tensor field equations 
\begin{equation}
2\nabla_{\alpha}\left(\sqrt{-g}\udt{\mathbb{P}}{\alpha}{\mu\nu}\right)+\sqrt{-g}
\left(\frac{\partial f}{\partial 
g^{\mu\nu}}-\frac{1}{2}fg_{\mu\nu}\right)=\frac{\sqrt{-g}}{\sigma}\,T_{\mu\nu}\,,
\label{STG_field_eqn}
\end{equation}
where 
\begin{equation}
T_{\mu\nu}:=-\frac{2}{\sqrt{-g}}\frac{\delta(\sqrt{-g}\mathcal{L}_{m})}{\delta 
g^{\mu\nu}}
\end{equation}
is the regular nonminimally coupled energy-momentum tensor for matter. Following the Palatini approach, the action can also be varied with respect to the connection, which results to the connection field equations
\cite{BeltranJimenez:2018vdo} 
\begin{equation}
\nabla_{\rho}\dut{\lambda}{\alpha}{\nu\mu\rho}+\dut{\lambda}{\alpha}{\mu\nu}=\sqrt{-g}\udt
{P}{\mu\nu}{\alpha}+\dut{H}{\alpha}{\mu\nu}\,,
\end{equation}
where 
\begin{equation}
\dut{H}{\alpha}{\mu\nu}:=-\frac{1}{2}\frac{\delta(\sqrt{-g}\mathcal{L}_{m})}{\delta\udt{
\Gamma}{\alpha}{\mu\nu}}
\end{equation}
is the hyper-momentum. The Lagrange multipliers can be eliminated through symmetry considerations to give the relation \cite{BeltranJimenez:2018vdo}
\begin{equation}
\mathbb{V}_{\alpha}:=\nabla_{\mu}\nabla_{\nu}\left(\sqrt{-g}\udt{\mathbb{P}}{\mu\nu}{
\alpha}\right)=0\,,\label{connection_field_eq}
\end{equation}
where it is assumed that $\nabla_{\mu}\nabla_{\nu}\dut{H}{\alpha}{\mu\nu}$ vanishes. It is possible to check that the connection equation \eqref{connection_field_eq} is trivially satisfied for the STEGR case \eqref{STEGR_Lagran}, since in this case it reduces to the
Bianchi identity.

\section{Newer General Relativity and Conformal Transformations \label{sec_ngr}}

In this section we present the Newer General Relativity, we extract the fields equations, 
and we investigate its features such as the conformal transformations.

\subsection{Newer General Relativity}

From Eq.(\ref{STEGR_Lagran}) it is clear that the STEGR scalar is quadratic in the non-metricity tensor. In fact, the non-metricity
tensor forms at most five parity-preserving independent scalars, which form the  irreducible set of invariants at quadratic order for STG. The theory with the Lagrangian given by a general linear combination of these five quadratic invariants was named \textit{newer general relativity} \cite{BeltranJimenez:2017tkd}, and is the straightforward symmetric teleparallel analogy of the \textit{new general relativity} introduced by Hayashi and Nakano in 1979 \cite{Hayashi:1979qx}.

The five parity-preserving independent quadratic scalars involve the two independent traces
\begin{equation}
Q_{\mu}:=\dudt{Q}{\mu}{\alpha}{\alpha}\,,\quad\quad\bar{Q}_{\mu}:=\dudt{Q}{\alpha}{\alpha}
{\mu}\,\label{inde_trace},
\end{equation}
which are used to form the set of quadratic irreducibles of STG, which
are $A_{1}:=Q_{\alpha\mu\nu}Q^{\alpha\mu\nu}$, $A_{2}:=Q_{\alpha\mu\nu}Q^{\mu\alpha\nu}$,
$A_{3}:=Q_{\alpha}Q^{\alpha}$, $A_{4}:=\bar{Q}_{\alpha}\bar{Q}^{\alpha}$,
$A_{5}:=\bar{Q}_{\alpha}Q^{\alpha}$. Together these scalar invariants
form the most general quadratic Lagrangian density known as newer GR, which is obtained 
by using $f = Q$ in (\ref{Gen_func_Lagran}), with 
\begin{eqnarray}
&&\!\!\!\!\!\!\!\!\!\!
Q:=c_{1}Q_{\alpha\mu\nu}Q^{\alpha\mu\nu}+c_{2}Q_{\alpha\mu\nu}Q^{\mu\alpha\nu}+c_{3}Q_{
\alpha}Q^{\alpha}\nonumber\\
&& \ \ +c_{4}\bar{Q}_{\alpha}\bar{Q}^{\alpha}+c_{5}\bar{Q}_{\alpha}Q^
{ \alpha } 
.\label{eq:Q}
\end{eqnarray}
This can be shown to produce STEGR Lagrangian in (\ref{STEGR_Lagran}) for
the coefficient choice \cite{BeltranJimenez:2018vdo} 
\begin{equation}
\mathcal{Q}=\frac{1}{4}Q_{\alpha\mu\nu}Q^{\alpha\mu\nu}-\frac{1}{2}Q_{\alpha\mu\nu}Q^{
\mu\alpha\nu}-\frac{1}{4}Q_{\alpha}Q^{\alpha}+\frac{1}{2}\bar{Q}_{\alpha}Q^{\alpha}\,,
\label{eq: Q GR}
\end{equation}
where the boundary term first introduced in (\ref{boundary_ein_def})
is then represented as 
\begin{equation}
B=\nabbol_{\alpha}\left(\bar{Q}^{\alpha}-{Q}^{\alpha}\right).
\end{equation}

Furthermore, the non-metricity conjugate for relation (\ref{eq:Q}), given by 
 (\ref{conj_Lag_ten}),
turns out to be 
\begin{align}
P^{\alpha}{}_{\mu\nu} & 
=c_{1}Q^{\alpha}{}_{\mu\nu}+c_{2}Q_{(\mu}{}^{\alpha}{}_{\nu)}+c_{3}g_{\mu\nu}Q^{\alpha}
\nonumber \\
 & \quad\, +
c_{4}\delta{}^{\alpha}{}_{(\mu}\widetilde{Q}_{\nu)}+\frac{c_{5}}{2}\left[\widetilde{Q}^{
\alpha}g_{\mu\nu}+\delta{}^{\alpha}{}_{(\mu}Q_{\nu)}\right]\,.\label{eq:P for Q}
\end{align}
In light of this non-metricity conjugate, the resulting metric field equations, by denoting the LHS as the tensorial density $E_{\mu\nu}$ for later convenience, are then given by
\begin{equation}
E_{\mu\nu}:=2\nabla_{\alpha}\left(\sqrt{-g}P^{\alpha}{}_{\mu\nu}\right)-q_{\mu\nu}-\frac{
\sqrt{-g}Q}{2}g_{\mu\nu}=\frac{\sqrt{-g}}{\sigma}\, T_{\mu\nu}\,,\label{newer_GR_FEs}
\end{equation}
which are easily derived using Eq.~(\ref{STG_field_eqn}), where $q_{\mu\nu}:=\frac{\partial\left(\sqrt{-g}Q\right)}{\partial 
g^{\mu\nu}}-\frac{1}{2}\sqrt{-g}Qg_{\mu\nu}$, and explicitly
\begin{align}
\frac{1}{\sqrt{-g}}q_{\mu\nu} & 
=c_{1}\left(2Q_{\alpha\beta\mu}Q^{\alpha\beta}{}_{\nu}-Q_{\mu\alpha\beta}Q^{\nu}{}_{
\alpha\beta}\right)\nonumber \\
 & \ \ \ +
c_{2}Q_{\alpha\beta\mu}Q^{\beta\alpha}{}_{\nu}+c_{3}\left(2Q_{\alpha}Q^{\alpha}{}_{\mu\nu}
-Q_{\mu}Q_{\nu}\right)\nonumber \\
 & \ \ \ +
c_{4}\widetilde{Q}_{\mu}\widetilde{Q}_{\nu}+c_{5}\widetilde{Q}_{\mu}Q^{\alpha}{}_{\mu\nu}
.\label{eq: q=0003BC=0003BD}
\end{align}

In the following subsections we will study the conformal transformations on the newer general relativity Lagrangian. For any Lagrangian of this class the important differences emerge at the level of the coefficients of the invariants. Hence, instead of writing Eq.~(\ref{eq:Q}) it proves convenient to introduce the following notation:
{\small{
\begin{align}
Q & =\left(\!\begin{array}{ccccc}
Q_{\alpha\mu\nu}Q^{\alpha\mu\nu} & Q_{\alpha\mu\nu}Q^{\mu\alpha\nu} & Q_{\alpha}Q^{\alpha} 
& \bar{Q}_{\alpha}\bar{Q}^{\alpha} & 
\bar{Q}_{\alpha}Q^{\alpha}\end{array}\!\right)\!\times\!\left(\!\begin{array}{c}
c_{1}\\
c_{2}\\
c_{3}\\
c_{4}\\
c_{5}
\end{array}\!\right),\label{eq:Q matrix form}
\end{align}}}
and to use the $\leftrightarrow$ to rewrite this as 
\begin{equation}
Q\leftrightarrow\left(\begin{array}{c}
c_{1}\\
c_{2}\\
c_{3}\\
c_{4}\\
c_{5}
\end{array}\right),
\end{equation}
where the irreducible scalar invariant basis is suppressed for readability. For instance, STEGR Lagrangian can then be represented as 
\begin{equation}
\mathcal{Q}\leftrightarrow\left(\begin{array}{c}
1/4\\
-1/2\\
-1/4\\
0\\
1/2
\end{array}\right)\,,
\end{equation}
instead of expression (\ref{eq: Q GR}), which will significantly reduce calculational complexity later on.

\subsection{Field Equations for a Quadratic Lagrangian\label{subsec:Field-Equations-for}}

Let us consider now a Lagrangian density that is quadratic in the linear combination of non-metricity invariants and hence quartic in the non-metricity tensor. Such a Lagrangian, as we argue later, appears naturally when formulating conformally invariant theory of gravity. Hence, we consider a quadratic Lagrangian of the form
\begin{eqnarray}
&&\!\!\!\!\!\!\!\!\!\!\!\!\!\!\!\!\!
\mathcal{L}_q=\frac{\sigma}{2}\sqrt{-g}\, f=\frac{\sigma}{2}\sqrt{-g}\, 
Q^{2}\nonumber\\
&&\!\!\!\!\!\!\!\!\!
=\frac{\sigma}{2} 
\sqrt{-g}\,(c_{1}A_{1}+c_{2}A_{2}+c_{3}A_{3}+c_{4}A_{4}+c_{5}A_{5})^{2},
\label{gen_quad_f}
\end{eqnarray}
and therefore the non-metricity conjugate resulting from expression (\ref{conj_Lag_ten})
then takes the form of
\begin{eqnarray}
&&\!\!\!\!\!\!\!\!\!\!\!\!\!\!\!\!\!\!\!
P'^{\alpha}{}_{\mu\nu} = 2 Q\times
\biggl\{c_{1}Q^{\alpha}{}_{\mu\nu}+c_{2}Q_{(\mu}{}^{\alpha}{}_{\nu)}+c_{3}g_{\mu\nu}Q^
{\alpha}\nonumber \\
 && \quad\quad \ \ \ 
c_{4}\delta{}^{\alpha}{}_{(\mu}\widetilde{Q}_{\nu)}+\frac{c_{5}}{2}\left[\widetilde{Q}^{
\alpha}g_{\mu\nu}+\delta{}^{\alpha}{}_{(\mu}Q_{\nu)}\right]\biggl\},
\label{eq:P for Q^2}
\end{eqnarray}
where it can be noted that
\begin{equation}
P'^{\alpha}{}_{\mu\nu}=2 QP^{\alpha}{}_{\mu\nu}\,.\label{eq:P' and P}
\end{equation}

The metric field equations are determined by inserting Eq.~(\ref{eq:P for Q^2}) into Eq.~(\ref{STG_field_eqn}), namely
\begin{eqnarray}
&&\!\!\!\!\!\!
2\nabla_{\alpha}\left(2\sqrt{-g} 
QP^{\alpha}{}_{\mu\nu}\right)-\sqrt{-g}\left[\frac{\partial\left( Q^{2}\right)}{\partial 
g^{\mu\nu}}-\frac{1}{2}\left( Q^{2}\right)g_{\mu\nu}\right]
\nonumber\\
&&\!\!\!\!\!\!
=\frac{\sqrt{-g}}{\sigma} \, 
T_{\mu\nu}\,,
\end{eqnarray}
which are the general field equations for the very general choice of $f$ in (\ref{gen_quad_f}). These field equations can be more compactly written as
\begin{align}
2QE_{\mu\nu}+t_{\mu\nu} & =\frac{\sqrt{-g}}{\sigma}T_{\mu\nu}\,,\label{eq:Q^2 EQM}
\end{align}
where $E_{\mu\nu}$ is defined in Eq. (\ref{newer_GR_FEs}) and $t_{\mu\nu}$ is defined as
\begin{align}
t_{\mu\nu} & 
:=\sqrt{-g}Q^{2}\left[\frac{-g_{\mu\nu}}{2}+4P^{\alpha}{}_{\mu\nu}\partial_{\alpha}\ln 
Q\right]\label{eq:t=0003BC=0003BD}.
\end{align}
Finally, the field equations for the connection can be calculated from \eqref{connection_field_eq}, since in this case $\mathbb{P}^{\alpha}{}_{\mu\nu}\equiv P'^{\alpha}{}_{\mu\nu}=2\sigma Q P^{\alpha}{}_{\mu\nu}$, and therefore they read as
\begin{equation}
V_{\alpha}:=\nabla_{\mu}\nabla_{\nu}\left(\sqrt{-g}QP^{\mu\nu}{}_\alpha\right)=0\,.
\label{eq: connection eq for Q2}
\end{equation}

\subsection{Conformal Transformations}

The fundamental dynamical object of STG is the metric tensor $g_{\mu\nu}$, which can undergo conformal transformations by means of the mapping
\begin{equation}
\widetilde{g}_{\mu\nu}\rightarrow\Omega^{2}(x)g_{\mu\nu}\,,\quad\quad\widetilde{g}^{\mu\nu
}\rightarrow\Omega^{-2}(x)g^{\mu\nu}\,,\label{Conf_trans_def}
\end{equation}
which represents local stretching, and where $\Omega^{2}(x)$ represents an arbitrary scalar field that has a dependence on manifold position. This so-called conformal factor depends on manifold spacetime coordinates $x^{\mu}$, which will be suppressed in what follows for brevity's sake.

As it is known, the above transformations preserve the causal structure of a manifold spacetime, and thus theories conformally related are dynamically equivalent. The effect on the metric determinant is then
\begin{equation}
\widetilde{g}=\Omega^{8}g\,,\label{metric_det_conf_trans}
\end{equation}
while the effect on the non-metricity tensor becomes
\begin{equation}
\widetilde{Q}_{\alpha\mu\nu}=\Omega^{2}\left(Q_{\alpha\mu\nu}+2g_{\mu\nu}\partial_{\alpha}
\ln\Omega\right)\,,
\end{equation}
with the associated inverse 
\begin{equation}
\widetilde{Q}^{\alpha\mu\nu}=\Omega^{-4}\left(Q^{\alpha\mu\nu}+2g^{\alpha\alpha'}g^{\mu\nu
}\partial_{\alpha'}\ln\Omega\right)\,.
\end{equation}
For the full set of irreducible scalar invariants we additionally need the conformal transformations of the two independent trace quantities given in Eq.(\ref{inde_trace}), which read as
\begin{align}
\widetilde{Q}_{\mu} & =Q_{\mu}+8\partial_{\mu}\ln\Omega\,,\\
\widetilde{\bar{Q}}_{\mu} & =\bar{Q}_{\mu}+2\partial_{\mu}\ln\Omega\,.
\end{align}
Combining these constituent terms gives the set of irreducibles of STG as 
\begin{eqnarray}
&&\widetilde{A}_{1} 
=\widetilde{Q}_{\alpha\mu\nu}\widetilde{Q}^{\alpha\mu\nu}\nonumber\\
&&\ \ \ \ \,
=
\Omega^{-2}
\Big[Q_{\alpha\mu\nu
}Q^{\alpha\mu\nu}+4Q^{\alpha}(\partial_{\alpha}\ln\Omega)
\nonumber\\
&&
 \ \ \ \ \ \ \ \ \ \ \ \ \ \ \,
+16g^{\alpha\alpha'}
(\partial_ {
\alpha}\ln\Omega)(\partial_{\alpha'}\ln\Omega)\Big]\,,
\label{Conf_irr_1}
\\
&&
\widetilde{A}_{2} 
=\widetilde{Q}_{\alpha\mu\nu}\widetilde{Q}^{\mu\alpha\nu}
\nonumber\\
&&\ \ \ \ \,
=\Omega^{-2}\Big[Q_{\alpha\mu\nu
}Q^{\mu\alpha\nu}+4\bar{Q}^{\alpha}(\partial_{\alpha}\ln\Omega)
\nonumber\\
&&
 \ \ \ \ \ \ \ \ \ \ \ \ \ \ \,
 +4g^{\alpha\alpha'}
(\partial_{\alpha}\ln\Omega)(\partial_{\alpha'}\ln\Omega)\Big]\,,
\label{Conf_irr_2}\\
&&
\widetilde{A}_{3}
=\widetilde{Q}_{\alpha}\widetilde{Q}^{\alpha}\nonumber\\
&&\ \ \ \ \,
=\Omega^{-2}\Big[Q_{\alpha}Q^{\alpha}+16Q^{
\alpha}(\partial_{\alpha}\ln\Omega)
\nonumber\\
&&
 \ \ \ \ \ \ \ \ \ \ \ \ \ \ \,
 +64g^{\alpha\alpha'}(\partial_{\alpha}
\ln\Omega)(\partial_{\alpha'}\ln\Omega)\Big]\,,
\label{Conf_irr_3}
\\
&&\widetilde{A}_{4} 
=\widetilde{\bar{Q}}_{\alpha}\widetilde{\bar{Q}}^{\alpha}\nonumber\\
&&\ \ \ \ \,
=\Omega^{-2}\Big[\bar{Q}_{\alpha
}\bar{Q}^{\alpha}+4\bar{Q}^{\alpha}(\partial_{\alpha}\ln\Omega)
\nonumber\\
&&
 \ \ \ \ \ \ \ \ \ \ \ \ \ \ \,
 +4g^{\alpha\alpha'}
(\partial_{\alpha}\ln\Omega)(\partial_{\alpha'}\ln\Omega)\Big]\,,
\label{Conf_irr_4}
\\
&&\widetilde{A}_{5}
=\widetilde{\bar{Q}}_{\alpha}\widetilde{Q}^{\alpha}
\nonumber\\
&&\ \ \ \ \,
=\Omega^{-2}\Big[\bar{Q}_{\alpha}Q^{
\alpha}+8\bar{Q}^{\alpha}(\partial_{\alpha}\ln\Omega)+2Q^{\alpha}(\partial_{\alpha}
\ln\Omega)
\nonumber\\
&&
 \ \ \ \ \ \ \ \ \ \ \ \ \ \ 
\,
+16g^{\alpha\alpha'}(\partial_{\alpha}\ln\Omega)(\partial_{\alpha'}
\ln\Omega)\Big]\label{Conf_irr_5}.
\end{eqnarray}

In order to give a simple example of these conformally transformed irreducible contributors, we consider the STEGR and boundary quantities which transform as
\begin{align}
\mathcal{Q} & 
=\Omega^{2}\widetilde{\mathcal{Q}}-2\Omega\left(\widetilde{\bar{Q}}^{\alpha}-\widetilde{Q}
^{\alpha}\right)\widetilde{\partial}_{a}\Omega-6\widetilde{\partial}^{\alpha}
\Omega\widetilde{\partial}_{\alpha}\Omega\\
B & 
=\Omega^{2}\widetilde{B}-2\Omega\left(\widetilde{\bar{Q}}^{\alpha}-\widetilde{Q}^{\alpha}
\right)\widetilde{\partial}_{a}\Omega-18\widetilde{\partial}^{\alpha}\Omega\widetilde{
\partial}_{\alpha}\Omega+6\Omega\widetilde{\Box}\Omega,
\end{align}
where 
$\widetilde{\Box}:=\widetilde{g}^{\alpha\beta}\widetilde{\nabbol}_{\alpha}\widetilde{\nabbol}_{\beta}$, and the covariant derivatives have been calculated with respect to the conformally transformed metric. This can be used to produce the Ricci scalar and its conformally transformed analogue, through the relation in (\ref{EH}), giving 
\begin{align}
\Rbol & =-\mathcal{Q}+B\nonumber \\ & 
=\Omega^{2}\left(-\widetilde{\mathcal{Q}}+\widetilde{B}\right)-12g^{\alpha\alpha'}
(\widetilde{\partial}_{\alpha}\Omega)(\widetilde{\partial}_{\alpha'}
\Omega)+6\Omega\widetilde{\Box}\Omega\nonumber \\ & 
=\widetilde{\Rbol}-12g^{\alpha\alpha'}(\widetilde{\partial}_{\alpha}\Omega)(\widetilde{
\partial}_{\alpha'}\Omega)+6\Omega\widetilde{\Box}\Omega\,,\label{eq:R to R conf}
\end{align}
which is precisely the same conformal transformation with that in the regular Levi-Civita determination of the Ricci scalar 
\cite{Faraoni:1998qx,ortin2007gravity},
as expected. Finally, for completeness, in Appendix \ref{Conf_irr_app} we present the inverse conformal transformations.

\section{Conformal Symmetric Teleparallel Gravity \label{sec_geoconf}}

In this section we introduce the purely geometric version of the conformal symmetric teleparallel gravity Lagrangian, built using the non-metricity tensor.

\subsection{The Conformal Lagrangian}

The conformal invariance at the level of the Lagrangian can be also considered through the newer GR Lagrangian introduced in (\ref{eq:Q}). This will produce a conformally invariant theory with conformally invariant field equations. However, as mentioned in \cite{Maluf:2011kf}, we must be very careful with forming this Lagrangian due to the conformal transformation of the volume element in the action. As shown in relation (\ref{metric_det_conf_trans}), the metric determinant transforms as $\widetilde{g}=\Omega^{8}g$, and therefore we must construct a Lagrangian with the form
\begin{equation}
\mathcal{L}_{c}=\frac{\sigma}{2} \sqrt{-g} Q^2\,,\label{Lagran_conf_cond}
\end{equation}
in order to cancel the conformal factor (in general in the above Lagrangian we could use a product of two distinct non-metricity scalars, namely $Q_1Q_2$, but for simplicity we choose to set them as the newer general relativity scalar $Q_{1}=Q_{2}:=Q$ defined in (\ref{eq:Q})). We can now exploit the condition that we need, namely
\begin{equation}
\widetilde{Q}=\Omega^{-2}Q\,,\label{eq: Sem Conf inv}
\end{equation}
which is equivalent to solving the system of equations 
\begin{equation}
\left(\begin{array}{ccccc}
4 & 0 & 16 & 0 & 2\\
0 & 4 & 0 & 4 & 8\\
16 & 4 & 64 & 4 & 16
\end{array}\right)\left(\begin{array}{c}
c_{1}\\
c_{2}\\
c_{3}\\
c_{4}\\
c_{5}
\end{array}\right)=0\,.\label{eq:conf syst}
\end{equation}
This is the result of using the conformally transformed components of the quadratic irreducible scalars of (\ref{Conf_irr_1})-(\ref{Conf_irr_5}), and substituting in the new GR Lagrangian (\ref{eq:Q}) so that the conformal condition (\ref{Lagran_conf_cond}) can be used for each of the constituents of the Lagrangian. This results in three independent equations which provide the general solution in terms of the three parameters $b_1,b_2,b_3\in\mathbb{R}$ as
\begin{equation}
\left(\begin{array}{c}
c_{1}\\
c_{2}\\
c_{3}\\
c_{4}\\
c_{5}
\end{array}\right)=\left(\begin{array}{c}
-b_1-4b_3\\
-4b_1-b_2\\
b_3\\
b_2\\
2b_1
\end{array}\right)\,.\label{eq:A_B_C}
\end{equation}
Note that due to the conformal invariance of the theory we can always absorb one of these factors to the other two, remaining with two free parameters. Hence, the Lagrangian (\ref{Lagran_conf_cond}) turns out to be 
\begin{equation}
\mathcal{L}_{c}=\frac{\sigma}{2} \sqrt{-g} Q_{c}^{2}\,,\label{eq:Geom Conf Lagr}
\end{equation}
where $Q_{c}$ is the general newer GR Lagrangian presented in (\ref{eq:Q}), for the choice of conformal parameters that satisfy the conditions (\ref{eq:A_B_C}), namely
\begin{eqnarray}
 &&
 Q_{c}^{2}= 
\big[-\left(b_1+4b_3\right)A_{1}-\left(4b_1+b_2\right)A_{2}
\nonumber\\
&&\ \ \ \ \ \ \ \ \ \,
+ b_3A_{3}
+ b_2A_{4} + 2b_1 A_{5}\big]{}^{2}\,.\label{gen_quad_f-1}
\end{eqnarray}

\subsection{Conformal Field Equations}

The conformal field equations are just a subclass of those derived for the $Q^{2}$ type of Lagrangians introduced in subsection \ref{subsec:Field-Equations-for}, but instead we use the conformal Lagrangian \eqref{eq:Geom Conf Lagr}. This effectively implies that the metric field equations are just
\begin{align}
2QE_{c\,\mu\nu}+t_{c\,\mu\nu} & =\frac{\sqrt{-g}}{\sigma}T_{\mu\nu}\,,\label{eq:Q^2 EQM-1}
\end{align}
where now $E_{c\, \mu\nu}$, $P_{c}^{\alpha}{}_{\mu\nu}$, $q_{c\,\mu\nu}$ and $t_{c\,\mu\nu}$ are just $P_{}^{\alpha}{}_{\mu\nu}$, $q_{\mu\nu}$ and $t_{\mu\nu}$ calculated in \eqref{eq:P for Q}, \eqref{eq: q=0003BC=0003BD} and \eqref{eq:t=0003BC=0003BD}, but evaluated with respect to the conformal coefficients
 \eqref{eq:A_B_C} as:
\begin{equation}
E_{c\,\mu\nu}=2\nabla_{\alpha}\left(\sqrt{-g}P_{c}^{\alpha}{}_{\mu\nu}\right)-q_{c\,
\mu\nu}-\frac{\sqrt{-g}}{2}Q_{c}\,g_{\mu\nu}\,,\label{Ec}
\end{equation}
\begin{align}
\!\!\!\!
P_{c}^{\alpha}{}_{\mu\nu} & 
=-\left(b_{1}\!+\!4b_{3}\right)Q^{\alpha}{}_{\mu\nu}-\left(4b_{1}\!+\!b_{2}\right)Q_{(\mu}
{} ^{
\alpha}{}_{\nu)}+b_{3}g_{\mu\nu}Q^{\alpha}\nonumber \\
 & 
\quad \, +
b_{2}\delta{}^{\alpha}{}_{(\mu}\widetilde{Q}_{\nu)}+
b_{1}
\left[\widetilde{Q}^{\alpha}g_{\mu\nu}+\delta{}^{\alpha}{}_{(\mu}Q_{\nu)}\right]\,,
\label{eq:P for Qc}
\end{align}
\begin{align}
\frac{1}{\sqrt{-g}}q_{c\,\mu\nu} & 
=-\left(b_{1}+4b_{3}\right)\left(2Q_{\alpha\beta\mu}Q^{\alpha\beta}{}_{\nu}-Q_{
\mu\alpha\beta}Q^{\nu}{}_{\alpha\beta}\right)\nonumber \\
 & 
\ \ \ \ 
+
b_{3}
\left(2Q_{\alpha}Q^{\alpha}{}_{\mu\nu}-Q_{\mu}Q_{\nu}\right)
+
b_{2}\widetilde{Q}_{\mu}\widetilde{Q}_{\nu}
\nonumber \\
 & 
\ \ \ \ 
-\left(4b_{1}+b_{2}\right)Q_{\alpha\beta\mu}Q^{\beta\alpha}{}_{\nu}+b_{1}
\widetilde{Q}_{\mu}Q^{\alpha}{}_{\mu\nu},\label{eq: q=0003BC=0003BD for Qc}
\end{align}
\begin{align}
t_{c\,\mu\nu} & 
:=\sqrt{-g}\,Q_{c}^{2}\left[\frac{-g_{\mu\nu}}{2}+4P_{c}^{\alpha}{}_{\mu\nu}\partial_{
\alpha}\ln Q_{c}\right]\,.\label{eq:t=0003BC=0003BD for Qc}
\end{align}

Finally, the field equations for the connection have already been calculated in \eqref{eq: connection eq for Q2}, and we only need to perform the appropriate substitutions as explained above for the metric field equations. Hence, we result to
\begin{equation}
\nabla_{\mu}\nabla_{\nu}\left(\sqrt{-g}Q_{c}P_{c}^{\alpha}{}_{\mu\nu}\right)=0.
\label{connection_field_eq-1}
\end{equation}

\subsection{Example: FLRW Metric \label{sec_example}}

%In the previous subsections we formulated %various conformal theories in the %symmetric teleparallel framework, and we %examined their relations through conformal %transformations. Hence, it would be %interesting to see what would be the %effects of these theories at a %cosmological setup.

We consider the spatially flat Friedmann-Lema\^itre-Robertson-Walker (FLRW) metric written in the proper time coordinate as 
\begin{equation}
ds^{2}=-dt^{2}+a(t)^{2}d\vec{x}^{2},\label{FRWLprop}
\end{equation}
where $a(t)$ is the cosmological scale factor and $d\vec{x}^{2}$ denotes the flat spatial element in the Cartesian coordinate system. As is known, in the case of Weyl Conformal Gravity \cite{Mannheim:2011ds} this is a trivial example since the corresponding Lagrangian  vanishes for the metric \eqref{FRWLprop}. This can be easily understood from the fact that when transforming the metric \eqref{FRWLprop} using the conformal time $d\eta=dt/a(t)$ it takes the form 
\begin{equation}
ds^{2}=a(\eta)^{2}\left(-d\eta^{2}+d\vec{x}^{2}\right),\label{FRWLconf}
\end{equation}
which is conformally equivalent to Minkowski spacetime. Since the conformally invariant theory cannot determine the conformal factor we deduce that the FLRW spacetime is a trivial example. The same holds for the Teleparallel Conformal Gravity \cite{Maluf:2011kf,Bamba:2013jqa}, where one finds that the conformally invariant teleparallel Lagrangian vanishes identically for the standard diagonal tetrad corresponding to the metric \eqref{FRWLprop}.

Having these in mind one could naively think that we should obtain the same result in the present case of conformally invariant symmetric teleparallel gravity. However, we encounter an interesting situation that illustrates an important feature concerning the applicability of the coincident gauge. In order to show this let us first consider the metric in \eqref{FRWLprop} and assume that it is in the coincident gauge, i.e. the corresponding symmetric teleparallel gravity is vanishing. In this case we find the quadratic non-metricity invariant as
\begin{equation}
Q=-12(c_{1}+3c_{3})H^{2}.
\end{equation}
Thus, for the choice of the parameters corresponding to the conformally invariant case \eqref{eq:A_B_C} we find that $Q_{c}=12(b_{1}+b_{3})H^{2}$, which does not vanish in general. This result can become more transparent if we examine the field equations for the connection and find that they are given by 
\begin{equation}
V_{\alpha}=\left\{ {432\,(b_1+b_3)\,a'\left(2a''^{2}+a'''a'\right)},0,0,0\right\} 
,\label{FRWLce}
\end{equation}
with primes denoting derivatives with respect to the conformal time. These equations are not generally satisfied, unless we impose the trivial constraint on the free parameters $b_1=-b_3$. Therefore, Eq.~\eqref{FRWLce} implies that the standard FLRW metric is not compatible with the coincident gauge. This is an interesting feature, since in the case of STEGR of \eqref{eq: Q GR} we find that the corresponding non-metricity invariant is $Q=6H^{2}$, i.e. is the same as in the ordinary Teleparallel Gravity. Hence, when we consider
the simplest $f(Q)$ extension we find that the corresponding connection equation is indeed satisfied for the metric \eqref{FRWLprop} and the coincident gauge, and thus $f(Q)$ is dynamically equivalent to the $f(T)$ gravity \cite{BeltranJimenez:2017tkd,Jimenez:2019ovq}.

This situation can be remedied in two ways. We either consider the non-vanishing symmetric teleparallel connection and find such a connection that the corresponding connection equations in Eq.~\eqref{connection_field_eq} are satisfied, or we search for other forms of the metric that correspond to the coincident gauge.

Let us state here that the first method seems to be rather hard to pursue. Despite the symmetric teleparallel connection having a rather simple pure gauge form in \eqref{conn} and being given by four independent function $\xi^{\mu}$ only, it is a rather difficult task
to analytically solve the connection equations. The reason for this is that the connection itself contains second derivatives of $\xi^{\mu}$ and the connection equations in Eq.~\eqref{connection_field_eq} are second order differential equations in connection. Therefore, we generally obtain fourth order differential equation for $\xi^{\mu}$, which are hard to solve analytically even in a such simple case as FLRW metric.

The much simpler method is to find the form of the metric that corresponds to the coincident gauge. We can show that the FLRW metric in the conformal time in \eqref{FRWLconf} is exactly such a choice since the non-metricity scalar is given by 
\begin{equation}
Q=-4\left(4c_{1}+c_{2}+16c_{3}+c_{4}+4c_{5}\right)\frac{\dot{a}^{2}}{a^{4}},
\end{equation}
where $\dot{a}=da/d\eta$.

We can immediately recognize the expression in the bracket to correspond to the bottom row of the matrix in \eqref{eq: Sem Conf inv} and
hence this expression being zero is one of the constraints given by the conformal invariance. Therefore, the conformally invariant Lagrangian \eqref{eq:Geom Conf Lagr} with Eq.~$\eqref{eq:A_B_C}$ identically vanishes 
\begin{equation}
Q_{c}=0.\label{FLRLlag}
\end{equation}

We can also check that the connection equations in Eq.~\eqref{connection_field_eq} are indeed satisfied for \eqref{FRWLconf} and zero connection
\begin{equation}
V_{\alpha}=\left\{ 0,0,0,0\right\} .\label{FRWLctc}
\end{equation}

Now, knowing that the conformal FLRW metric \eqref{FRWLconf} is in the coincident gauge, it is an easy task to generate connection for the FLRW metric in the proper time by simply transforming to the conformal time and showing that this coordinate change generates a non-vanishing symmetric teleparallel connection with only one non-vanishing component 
\begin{equation}
\Gamma^{0}{}_{00}=-H.\label{FLRWconn}
\end{equation}
We can then show that the FLRW metric in the proper time \eqref{FRWLprop} and the connection in \eqref{FLRWconn} indeed solve the connection equations in \eqref{connection_field_eq} and lead to the vanishing Lagrangian \eqref{FLRLlag}.

While the form of the connection \eqref{FLRWconn} is extraordinarily simple, it is not an easy task to obtain this result by considering
a general coordinate transformation and solving the corresponding fourth order differential equations. The presented method of guessing
the metric in the coincident gauge and generating the symmetric teleparallel connection by coordinate change is much simpler. Its obvious shortcoming is that it may be practically impossible to guess the metric in the coincident gauge in more complicated cases than FLRW.

Another interesting insight obtained from this simple example is that the coincident gauge depends on the chosen theory Lagrangian, which can be seen when we compare $f(Q)$ gravity and our conformal symmetric teleparallel gravity. Both theories have generally non-trivial field equations for the connection. However, in the case of the metric \eqref{FRWLprop} they are solved with vanishing connection only for $f(Q)$ gravity, while our conformal symmetric teleparallel gravity requires the non-vanishing connection \eqref{FLRWconn}. Therefore, the metric \eqref{FRWLprop} is in the coincident gauge in the case of $f(Q)$ gravity but not in the conformally invariant theory considered here.

While generally it is expected that the given metric is in the coincident gauge should depend on the underlying theory, it is rather unexpected that it happens in the FLRW case when compared to the situation in torsional Teleparallel Gravity. In torsional Teleparallel Gravity we face the similar issue whether--and if yes under what circumstances--to a given tetrad corresponds a vanishing teleparallel spin connection \cite{Krssak:2015rqa,Krssak:2015lba,Tamanini:2012hg,Krssak:2015oua,Golovnev:2017dox}. In the case of FLRW spacetimes, it was shown that for all modified teleparallel theories the tetrad has universally the same form as in Ref. \cite{Hohmann:2019nat}. This illustrates significant differences between torsional and non-metricity teleparallel theories.

Another feature to note is that while the flat FLRW metric is conformally flat, the Universe is only globally homogeneous. On local scales, such as clusters and superclusters, the Universe exhibits inhomogeneities that break this symmetry which may effect the dynamics of these structures as compared with standard gravity. Also, as noted in the case of Weyl gravity (Levi-Civita based Conformal Gravity) in Refs.\cite{Mannheim:2011ds,Mannheim:2005bfa}, the potential contribution of a very small deviation from purely flat cosmology would produce a non-vanishing effect in the conformal Friedmann equations. This instability in the Friedmann equations appears in GR too in the form of the flatness problem. It would be interesting to investigate further the effect of a slightly nonflat cosmology within this conformal expression of torsional Teleparallel Gravity.

\subsection{Conformal Invariance and a Scalar Field \label{sec_scalar}}

An alternative way of constructing a conformally invariant action from $Q_{c}$ is by including a free scalar field, similarly to \cite{Maluf:2011kf}. This scalar field is a conformal density, since it must also transform as $\widetilde{\varphi}=\Omega^{-1}\varphi$.
The general form of the action is then given by
\begin{eqnarray}
&&\!\!\!\!\!\!\!\!\!\!\!\!\!\!\!\!\!\!\!\!
S_{c}=\int 
d^{4}x\,\Big[\frac{\sigma}{2}\sqrt{-g}\,(-\varphi^{2}Q_{c}+g^{\mu\nu}\text{D}_{\mu}
\varphi 
D_{\nu}\varphi)
 \nonumber\\
&& \ \ \ \ \ \ \ 
+\dut{\lambda}{\alpha}{\beta\mu\nu}\udt{R}{\alpha}{\beta\mu\nu}+\dut{
\lambda}{\alpha}{\mu\nu}\udt{T}{\alpha}{\mu\nu}\Big]\,,
\label{eq:Conf Lagr 
=00003D00003D0003C6}
\end{eqnarray}
where $\text{D}_{\mu}:=\partial_{\mu}+A_{\mu}$ is the covariant derivative and $A_{\mu}$ is an one-form. This covariant derivative will be endowed with transformation properties from the kinetic term itself. The transformation law of the kinetic term must be
\begin{equation}
\widetilde{g}^{\mu\nu}\widetilde{\text{D}_{\mu}\varphi}\widetilde{\text{D}_{\nu}\varphi}
=\Omega^{-4}g^{\mu\nu}\text{D}_{\mu}\varphi \text{D}_{\nu}\varphi,
\end{equation}
to ensure conformal invariance of the action, then $\text{D}_{\mu}$ must be conformally covariant, i.e 
\begin{equation}
\widetilde{\text{D}_{\mu}\varphi} = 
\widetilde{\text{D}_{\mu}}\widetilde{\varphi}=\Omega^{-1}\text{D}_{\mu}\varphi,
\end{equation}
and thus $A_{\mu}$ should transform as 
\begin{equation}
\widetilde{A}_{\mu}=A_{\mu}+\partial_{\mu}\ln\Omega.\label{eq:A=00003D00003D0003BC}
\end{equation}
The most general $A_{\mu}$, constructed by non-metricity parts, that
satisfies (\ref{eq:A=00003D00003D0003BC}) is 
\begin{equation}
A_{\mu}=\alpha Q_{\mu}+\frac{1}{2}\left(1-8\alpha\right)\overline{Q}_{\mu},
\end{equation}
where $\alpha\in\mathbb{R}$. A very useful quantity is the the norm of 
$A_{\mu}$:
\begin{eqnarray}
&&\!\!\!\!\!\!\!\!\!\!\!\!\!\!\!\!\!\!\!\!\!\!\!
A_{\mu}A^{\mu}=\alpha^{2}Q_{\mu}Q^{\mu}+\alpha\left(1-8\alpha\right)Q^{\mu}\overline{Q}_{
\mu}\nonumber\\
&&+\left[\frac{1}{2}\left(1-8\alpha\right)\right]^{2}\overline{Q}_{\mu}\bar{Q
} ^ { \mu } ,
\label{eq:A=00003D00003D0003BC=00003D00003D000391=00003D00003D0003BC}
\end{eqnarray}
which can also be recasted in matrix form for further simplification as 
\begin{equation}
A_{\mu}A^{\mu}=\left(\begin{array}{c}
0\\
0\\
\alpha^{2}\\
\left[\frac{1}{2}\left(1-8\alpha\right)\right]^{2}\\
\alpha\left(1-8\alpha\right)
\end{array}\right).\label{eq:A=00003D00003D0003BC=00003D00003D000391=00003D00003D0003BC 
matrix}
\end{equation}
Notice that in order for the cross term $Q^{\mu}\overline{Q}_{\mu}$ to be present, both $Q_{\mu}Q^{\mu}$ and $\overline{Q}_{\mu}\bar{Q}^{\mu}$ need to appear too. In the case where $\alpha=1/8$ we have $A_{\mu}A^{\mu}=\left(1/64\right)Q_{\mu}Q^{\mu}$, and for $\alpha=0$ we have $A_{\mu}A^{\mu}=\left(1/4\right)\overline{Q}_{\mu}\bar{Q}^{\mu}$. This feature will be very important in finding whether or not we can have the non-conformal limit, as we see below. A similar analysis was performed in Ref. \cite{Iosifidis:2018zwo}, using a more general Lagrangian including torsion but a less general connection that is built from only one of the non-metricity traces of (\ref{inde_trace}).

The Lagrangian of (\ref{eq:Conf Lagr =00003D00003D0003C6}) should have a non-conformal limit to $\mathcal{Q}$ for some choice of $\alpha$ while $\varphi=const.(=1)$. In order to investigate this limit we split $Q_c$ into the STEGR part $\mathcal{Q}$ and the remaining $\Delta Q$ as
\begin{equation}
Q_{c}=\mathcal{Q}+\Delta Q.
\end{equation} 
The non-conformal limit will be obtained by demanding that $\Delta Q$ is completely canceled out by the kinetic term when $\varphi=const.$, while also having substituted the STEGR coefficients $c_{1}=-c_3=1/4,$ $c_{2}=-c_{5}=-1/2$ and $c_{4}=0$.

At this stage we have to choose a specific solution from (\ref{eq:conf syst}) that will lead to the realization of the non-conformal limit. Considering the form (\ref{eq:A=00003D00003D0003BC=00003D00003D000391=00003D00003D0003BC}), for this to happen we have three possible choices of the real free parameters
\begin{itemize}
\item Case 1: \label{enu:,c1,c2,c3} $\left(\begin{array}{c}
c_{1}\\
c_{2}\\
c_{3}\\
c_{4}\\
c_{5}
\end{array}\right)=\left(\begin{array}{c}
c_{1}\\
c_{2}\\
c_{3}\\
4c_{1}-c_{2}+16c_{3}\\
-2c_{1}-8c_{3}
\end{array}\right),$ 
\item Case 2: \label{enu:,c1,c2,c4} $\left(\begin{array}{c}
c_{1}\\
c_{2}\\
c_{3}\\
c_{4}\\
c_{5}
\end{array}\right)=\left(\begin{array}{c}
c_{1}\\
c_{2}\\
\frac{1}{16}\left(-4c_{1}+c_{2}+c_{4}\right)\\
c_{4}\\
-\frac{1}{2}\left(c_{2}+c_{4}\right)
\end{array}\right),$ 
\item \label{enu:,c1,c2,c5} Case 3: $\left(\begin{array}{c}
c_{1}\\
c_{2}\\
c_{3}\\
c_{4}\\
c_{5}
\end{array}\right)=\left(\begin{array}{c}
c_{1}\\
c_{2}\\
\frac{1}{8}\left(-2c_{1}-c_{5}\right)\\
-c_{2}-2c_{5}\\
c_{5}
\end{array}\right).$ 
\end{itemize}
Hence, 
calculating $\Delta Q$ directly in the non-conformal limit for each case we obtain
\begin{itemize}
\item Case 1: $\Delta Q=\left(\begin{array}{c}
0\\
0\\
0\\
-5/2\\
1
\end{array}\right),$ 
\item Case 2: $\Delta Q=\left(\begin{array}{c}
0\\
0\\
5/32\\
0\\
-1/4
\end{array}\right),$ 
\item Case 3: $\Delta Q=\left(\begin{array}{c}
0\\
0\\
1/8\\
-1/2\\
0
\end{array}\right)\,.$ 
\end{itemize}
Notice that in each case we have two non-zero components that must be canceled out from the kinetic term in the non-conformal limit. By re-writing the kinetic term in matrix notation we have 
\begin{equation}
g^{\mu\nu}\text{D}_{\mu}\varphi 
\text{D}_{\nu}\varphi{\rightarrow}A_{\mu}A^{\mu}=\left(\begin{array}{c}
0\\
0\\
\alpha^{2}\\
\left[\frac{1}{2}\left(1-8\alpha\right)\right]^{2}\\
\alpha\left(1-8\alpha\right)
\end{array}\right)\,.\label{eq:kinetic term N-CL}
\end{equation}
Therefore, the correct choice of $\Delta Q$ is the one that satisfies
\begin{equation}
-\Delta Q+A_{\mu}A^{\mu}=0\,,\label{eq:N-CL cond}
\end{equation}
for some of the three available choices of parameteres in 
Cases 1,2,3 above.

Apparently, none of our available choices in conjunction with an appropriate choice of $\alpha$ satisfies Eq.~(\ref{eq:N-CL cond}), and therefore we cannot have such a physically interesting non-conformal limit, although the Lagrangian is conformally invariant. The fact that the theory has no connection to the non-conformal limit, and one has to include the extra parameter $\alpha$, it is a disadvantage, in contrast to the geometrical one introduced in Sec.~\ref{sec_geoconf}. This is a notable difference between the non-metricity conformal teleparallel gravity and torsional conformal teleparallel gravity \cite{Maluf:2011kf}, where a proper non-conformal limit was found, and the apparent reason for that lies solely on the difference between the non-metricity and torsion tensors. This relationship requires further investigation.

\section{Scalar-Tensor Form of \texorpdfstring{$f(\mathcal{Q},B)$}{f(Q,B)} gravity \label{sec_fqb}}

In this section we present the scalar-tensor form of $f(\mathcal{Q})$ and $f(\mathcal{Q},B)$ gravities, and we investigate their coupling with a scalar field. In particular, the scalar-tensor representation of a theory, in conjunction with the Einstein Frame, is a powerful tool that can provide further insight by reformulating the theory itself in an equivalent way by introducing extra scalar field(s). In this way, one obtains a different, but equivalent, way of probing the propagating degrees of freedom of the theory.

\subsection{Scalar-Tensor Form of \texorpdfstring{$f(Q)$}{f(Q)} gravity}

We first start by investigating the scalar-tensor representation of $f(\mathcal{Q})$ theory and examining its relation to torsional teleparallel theories. We apply the usual scalar-tensor representation method by introducing two auxiliary fields $\chi$
and $\varphi$ and we write
\begin{eqnarray}
&&\!\!\!\!\!\!\!\!\!
S=\int d^4x\sqrt{-g}\,\frac{\sigma}{2}f(\mathcal{Q})dx\nonumber\\
&&\!\!\!
=\int 
d^4x\sqrt{-g}\,\frac{\sigma}{2}\left[
\chi(\mathcal{Q}-\varphi)+f(\varphi)\right],
\end{eqnarray}
which can be further recasted by choosing $\chi=f'(\varphi):=-F(\varphi)$ as 
\begin{equation}
S=\int d^4x\sqrt{-g}\,\frac{\sigma}{2}\left[-F(\varphi)\mathcal{Q}-V(\varphi)\right]\,,
\end{equation}
where $V(\varphi):=f(\varphi)-\varphi f'(\varphi)$. Hence, this can be transformed to the Einstein frame as 
\begin{align}
S &= \int d^4x\sqrt{-\widetilde{g}}\, \Omega^{-4} \frac{\sigma}{2} \,\Biggl\{-F(\varphi) \Big[\Omega^{2}\widetilde{\mathcal{Q}}\!
\nonumber\\
& -2\Omega\left(\widetilde{\bar{Q}}^{\alpha}\!-\!\widetilde{Q}^{\alpha}\right)\widetilde{\partial}_{a}\Omega-6\widetilde{\partial}^{\alpha}\Omega\widetilde{\partial}_{\alpha} \Omega\Big]-V(\varphi)\Biggl\}.
\end{align}
Using integration by parts, we finally result to
\begin{align}
S &= \int d^4x\sqrt{-\widetilde{g}}\,\frac{\sigma}{2} \Big[-\widetilde{\mathcal{Q}}-\frac{1}{\sqrt{3}}\psi\widetilde{B}+\frac{1}{2}\widetilde{\nabla_{\alpha}}\psi\widetilde{\nabla^{\alpha}}
\psi \nonumber\\
& - U(\psi)\Big]\,,\label{eq:Scalar ten f(Q)}
\end{align}
where $\psi:=\sqrt{3}\ln F(\varphi)$, $\widetilde{B}:=\mathcal{\widetilde{D}}_{\alpha}\left(\widetilde{\overline{Q}}^{\alpha}-\widetilde{Q}^{\alpha}\right)$ and $U(\varphi):={\displaystyle \frac{V(\varphi)}{F(\varphi)^{2}}}$. This action corresponds to the regular STEGR theory plus a scalar field $\psi$ coupled to the boundary term in the Einstein frame. Remarkably it has the exact same functional form as in the torsional teleparallel gravity case presented in \cite{Wright:2016ayu}. This also implies that it suffers from the incorrect sign in the kinetic term of the scalar field $\psi$, which makes it to be a phantom field.

\subsection{STG Dark Energy}

In curvature gravity one class of theories is obtained by introducing a scalar field,non-minimally coupled with the Ricci scalar. Similarly, in the case of teleparallel gravity one can introduce a scalar field and allow for a non-minimal coupling with the torsion scalar, in the class of theories called ``teleparallel dark energy'' \cite{Geng:2011aj,Xu:2012jf,Otalora:2013dsa,Hohmann:2018rwf}. In this subsection we follow the same approach in the framework of STEGR theory.

We start using the action \eqref{eq:Scalar ten f(Q)} as a template, and we slightly change the coupling $\psi B\rightarrow\xi\mathcal{Q}\psi$. Then, renaming $\psi\rightarrow\varphi$ we obtain
\begin{equation}
S=\int d^4x \sqrt{-g}\, \frac{\sigma}{2} \left[-{\displaystyle 
\frac{\mathcal{Q}}{2}}+\frac{1}{2}\nabla_{\alpha}\varphi\nabla^{\alpha}\varphi+\xi\mathcal
{Q}\varphi^{2}-V(\varphi)\right],\label{eq:Coin dark En}
\end{equation}
which can be further generalized by introducing an arbitrary coupling for $\mathcal{Q}$:
\begin{equation}
S=\int d^4x \sqrt{-g}\, 
\frac{\sigma}{2} \left[-A(\varphi)\mathcal{Q}+\frac{1}{2}\nabla_{\alpha}
\varphi\nabla^{\alpha}\varphi-V(\varphi)\right]\,\label{eq:Gen Coupl}
\end{equation}
(thus (\ref{eq:Coin dark En}) is obtained for the choice $A(\varphi)=\frac{1}{2}-\xi\varphi^{2}$).

As we can see, the theory (\ref{eq:Gen Coupl}) cannot be written as an $f(\mathcal{Q})$ theory of the form (\ref{eq:Scalar ten f(Q)}), since there is no coupling with the boundary term. Nevertheless, we can indeed bring it in the form of an $f(\mathcal{Q},B)$ theory. In particular, we transform to the Einstein frame as
\begin{eqnarray}
&&
\!\!\!\!\!\!\!\!\!\!\!\!\!\!\!\!
S = \int d^4x\sqrt{-\widetilde{g}}\, \Omega^{-4} \frac{\sigma}{2} \Biggl\{
\frac{1}{2}\Omega^{2}\widetilde{\nabla}^{\alpha}\varphi\widetilde{\nabla}_{\alpha}
\varphi-V(\varphi)
\nonumber \\
& & \!\!\!\!\!\!\!\!\!\!
-A(\varphi)\!
\left[\Omega^{2}\widetilde{\mathcal{Q}}-2\Omega\left(\widetilde{
\bar{Q}}^{\alpha}\!-\!\widetilde{Q}^{\alpha}\right)\widetilde{\partial}_{a}\Omega
 -6\widetilde{
\partial}^{\alpha}\Omega\widetilde{\partial}_{\alpha}\Omega\right]\!
 \Biggl\},\label{eq:S coin Einst}
\end{eqnarray}
and we eliminate the kinetic term
$\frac{1}{2}\Omega^{2}\widetilde{\nabla}^{\alpha}\varphi\widetilde{\nabla}_{\alpha} \varphi$ by demanding that $6\Omega^{-4}A(\varphi)\widetilde{\partial}^{\alpha}\Omega\widetilde{\partial}_{\alpha}\Omega = \frac{1}{2}\Omega^{-2}\widetilde{\nabla}^{\alpha}\varphi\widetilde{\nabla}_{\alpha}\varphi$, which for $A(\varphi)\neq0$ is satisfied by the following solution for $\Omega(\varphi)$:
\begin{equation}
\Omega(\varphi)=\exp\left[\int{\displaystyle 
\frac{d\varphi}{2\sqrt{3A(\varphi)}}}\right]\,.
\end{equation}
Hence, by substituting into (\ref{eq:S coin Einst}) and introducing the function $G(\Omega)=\int\frac{A(\Omega)}{\Omega^{3}}d\Omega$ we acquire
\begin{align}
S & = \int d^4x\sqrt{-\widetilde{g}}\, \Omega^{-4} \frac{\sigma}{2} \Big[-A(\Omega)\Omega^{-2}\widetilde{\mathcal {Q}}-G(\Omega)\widetilde{B} \nonumber\\
& - U(\Omega)\Big]\,,\label{eq:f(Q,B)}
\end{align}
where now $\Omega$ is an auxiliary field. Its field equation is extracted as
\begin{align}
\left[
\frac{2A(\Omega)-\Omega 
A'(\Omega)}{\Omega^{3}}\right]\widetilde{\mathcal{Q}}-\frac{A(\Omega)}{\Omega^{3}}
\widetilde { B }
-U'(\Omega) & =0,\label{eq: field eq =0003A9}
\end{align}
and since this is a first order ordinary differential equation there is always a solution and therefore we can extract $\Omega$ as
$\Omega=\Omega(\widetilde{\mathcal{Q}},\widetilde{B})$. Hence, we can finally re-write (\ref{eq:f(Q,B)}) as 
\begin{align}
S & 
=\int d^4x\sqrt{-\widetilde{g}}\, \frac{\sigma}{2}
\left[f(\widetilde{\mathcal{Q}},\widetilde{B}
)\right].
\label{eq:f(Q,B)2}
\end{align}

In summary, we showed that the interesting class of non-minimal STEGR can be re-written as a modified $f(\mathcal{Q},B)$ gravity. This property makes $f(\mathcal{Q},B)$ theories cosmologically interesting, which is also a shared property of $f(T,B)$ theories \cite{Wright:2016ayu}.

\subsection{Scalar-Tensor Form of \texorpdfstring{$f(Q,B)$}{f(Q,B)} gravity}

Following the usual procedure as in the previous subsections, we introduce auxiliary scalar fields and thus $f(\mathcal{Q},B)$ theory can be reformulated as
\begin{align}
S &= \int d^4x \sqrt{-g}\, \frac{\sigma}{2}\, f(\mathcal{Q},B)\nonumber \\
 & = \int d^4x \sqrt{-g}\, \frac{\sigma}{2}\Big[\chi_{1}(\mathcal{Q}-\varphi_{1})+\chi_{2}(B-\varphi_{
2}) \nonumber\\
& + f(\varphi_{1},\varphi_{2})\Big]\,,\label{eq:ST f(Q,B)}
\end{align}
where $\varphi_{1},\varphi_{2},\chi_{1},\chi_{2}$ are auxiliary fields
for which 
\begin{equation}
\begin{array}{ccccc}
{\displaystyle \frac{\delta S}{\delta\varphi_{1}}}=0\Rightarrow & 
\chi_{1}=\partial_{\varphi_1}f, & & 
{\displaystyle \frac{\delta S}{\delta\varphi_{2}}}=0\Rightarrow & 
\chi_{2}=\partial_{\varphi_2}f\,,\\
\\
{\displaystyle \frac{\delta S}{\delta\chi_{1}}}=0\Rightarrow & \varphi_{1}=\mathcal{Q}, &
& {\displaystyle \frac{\delta S}{\delta\chi_{2}}}=0\Rightarrow & B=\varphi_{2}\,,
\end{array}\label{eq:Aux var}
\end{equation}
with $\partial_{\varphi_i}f=\partial f/\partial\varphi_{i}$ and $\partial_{\varphi_i\varphi_i}f=\partial^2 f/\partial\varphi_{i}^2$. Assuming that $ \partial_{\varphi_1\varphi_1} f,\partial_{\varphi_2\varphi_2} f\neq0$ in order to have the full form of $f(\mathcal{Q},B)$, since 
$\partial_{\varphi_2\varphi_2} f=0$ would imply $f=f(\mathcal{Q})$, we finally obtain
\begin{align}
S\!& = \int \! d^4x\sqrt{-g}\,\frac{\sigma}{2}\Big[F(\varphi_{1},\varphi_{2})\mathcal{Q}+G(\varphi_{1},\varphi_{2})B \nonumber\\
& -V(\varphi_{1},\varphi_{2})\Big]\,.\label{eq:ST form of f(Q,B)}
\end{align}
Transforming again to the Einstein frame, and performing integration by parts,
we find
\begin{align}
S &= \int \!d^4 x \sqrt{-\widetilde{g}}\, \frac{\sigma}{2}\Big\{G\Big[\Omega^{-2}\widetilde{B}-2\Omega^{-3}\left(\widetilde{\bar{Q}}^{\alpha} - \widetilde{Q}^{\alpha}\right)\widetilde{\partial}_{a}\Omega\Big]\nonumber\\
 & + F\left[\Omega^{-2}\widetilde{\mathcal{Q}}-2\Omega^{-3}\left(\widetilde{\bar{Q}}^{\alpha}
\!-\!\widetilde{Q}^{\alpha}\right)\widetilde{\partial}_{a}\Omega-6\Omega^{-4}\widetilde{
\partial}^{\alpha}\Omega\widetilde{\partial}_{\alpha}\Omega\right] \nonumber \\
 & -6\Omega^{-3}\widetilde{\partial^{\mu}}\Omega\widetilde{\partial_{\mu}}G-U\Big\}\,,\label{eq:Einstein f(Q,B)-1}
\end{align}
where for simplicity we have suppressed the 
arguments in $F(\varphi_{1},\varphi_{2}),G(\varphi_{1},\varphi_{2})$ and 
$V(\varphi_{1},\varphi_{2})$. The coupling between $\widetilde{\mathcal{Q}}$ and $\varphi$ can be eliminated by making the choice of $F=-\Omega^{2}$. Moreover, we can eliminate the boundary term too, by integrating it out and then factorizing
all the terms contracted with $\left(\widetilde{\bar{Q}}^{\alpha}-\widetilde{Q}^{\alpha}\right)$ and setting them to zero. After some simplifications we result to $2F\Omega^{-1}\widetilde{\partial}_{\varphi_i}\Omega+\widetilde{\partial}_{\varphi_i}G=0
$, and therefore 
using the chain rule we obtain two equations, namely
\begin{eqnarray}
&&\partial_{\varphi_1}\Omega = -\Omega \partial_{\varphi_1}G/2F\,,\label{eq:coupl eq 
1-1}\\
&&\partial_{\varphi_2}\Omega = -\Omega \partial_{\varphi_2}G/2F\,.\label{eq:coupl eq 
1-2}
\end{eqnarray}
By demanding $\Omega(\varphi_{1},\varphi_{2})$ only to be smooth and continuous we can differentiate Eq.~(\ref{eq:coupl eq 1-1}) with respect to the second argument, then differentiate Eq.~(\ref{eq:coupl eq 1-2}) with respect to the first argument, and then by equating them and replacing $G$ and $F$ with $f$, we arrive at the new system of equations 
\begin{eqnarray}
\partial_{\varphi_2}\Omega \ \partial_{\varphi_2\varphi_1}f-\partial_{\varphi_1} \Omega \
\partial_{\varphi_2\varphi_2} f=0\,,
\label{eq:coupl eq 2-1}\\
\partial_{\varphi_2\varphi_2} f\ \partial_{\varphi_1\varphi_1}
f-(\partial_{\varphi_1\varphi_2} f)^{2}=0\,,\label{eq:coupl eq 2-2}
\end{eqnarray}
where a first straightforward solution is found to be $f(\varphi_{1},\varphi_{2})=\exp\left[\partial_{\varphi_1}\Omega\,\varphi_{1}+\partial_{
\varphi_2 } \Omega\,\varphi_{2} \right]$. With this in hand, any $f(\mathcal{Q},B)$ theory can be reformulated into a scalar-tensor form $A(\mathcal{\widetilde{Q}},\varphi_{1},\varphi_{2})$ for some function $A$.
%Note that 
% in \cite{Wright:2016ayu} Eq. (\ref{eq:coupl eq 2-1}) was also found but 
%does not comment explicitly
%on it, that an obvious solution is $\Omega=f_{1}$ which makes it identical
%to Eq.~(\ref{eq:coupl eq 2-2}). On the other hand, Eq.~(\ref{eq:coupl eq 2-2})
%is the same as the one presented in Ref.\cite{Wright:2016ayu}.

\section{Conclusion \label{sec_conclusion}}

In this article, we presented the case for considering conformal symmetries in STG through several different approaches. In Sec.~\ref{sec_ngr}, we consider the constituents of STG through the irreducible components of the non-metricity tensor. In turn, we present each of the conformally transformed elements that make up this theory, as well as the effect on the metric tensor. One result of this approach is that by using the STG constituents, we arrive at the same conformal relation in Eq.~(\ref{eq:R to R conf}) for the Ricci scalar as compared with GR. However, STG offers a much richer framework to work in since it separates the second and fourth (boundary) contributions of the Ricci scalar. This point permeates through to the extensions of the theory. To form a conformally invariant theory of gravity, the full Lagrangian of the theory must be conformally invariant, which is more than simply the Lagrangian density. In Sec.~\ref{sec_geoconf} this proposal is explored fully with a quartic order Lagrangian emerging as a requirement to counter-balance the contribution from the conformally transformed metric determinant. The result is a class of theories with two free coefficients presented in Eq.~(\ref{eq:A_B_C}) where one the three apparent coefficients being absorbed by the conformal invariance of the theory. In Eq.~(\ref{eq:Q^2 EQM-1}) we use this new Lagrangian to write explicitly the resulting field equations.

A well-known property of conformal theories of gravity is a triviality of homogenous cosmologies since the spatially flat FLRW metric is conformally flat. In subsection \ref{sec_example} we discuss this issue and ways that this metric can be considered within conformally symmetric teleparallel theory. Principally, we find that the flat FLRW metric in the standard Cartesian coordinates is not compatible with the coincident gauge within the conformal theory and so we either need to solve the connection field equations or find another form of the metric that is compatible. While the former choice may seem straightforward, the equations quickly become almost intractable to solve, and so we opt for the latter choice. Immediately, we find that the FLRW metric in conformal time solves the problem and we retrieve a vanishing Lagrangian for this instance. While the FLRW metric results in vanishing field equations, the real Universe contains inhomogeneities on the local scale that contribute non-negligibly to the gravitational potentials \cite{Mannheim:2011ds}. In this way, conformal theories may have some interesting features for globally conformal systems.

It would also be interesting to investigate the behavior of spherically symmetric solutions in this regime given the enormous success conformal theories have had in standard gravity (such as in providing analytic spherically symmetric solutions \cite{Mannheim:1988dj} and a potential resolution to the rotation curve problem for galaxies \cite{OBrien:2011vks}).

Another intriguing extension to modified gravity is that of a coupled scalar field. In subsection \ref{sec_scalar} we take the case of a
nonminimally coupled scalar field where the coincident gauge no longer applies. In this particular case, the connection can be solved analytically under the demand that the Lagrangian is conformally invariant. This results in three possible dependencies for the free parameters of the theory. However, none of these potential choices produce a sensible limit to STEGR in the limit of a constant scalar field which is dissimilar to the situation in Teleparallel Gravity where a physical limit is achieved \cite{Maluf:2011kf}.

In Sec.~\ref{sec_fqb}, we finally consider the $f(Q,B)$ extension to STEGR in which the second and fourth order contributions to $f(R)$ gravity are individually separated. Interestingly, in this case we find that this extension to STG can indeed be represented as a scalar-tensor theory. We also show that at least one dark energy scenario can be transformed to an $f(Q,B)$ model, which
is interesting since this extension has a much richer class of models comparing to the popular $f(R)$ gravity. It is interesting to compare our symmetric teleparallel conformal gravity with the standard Weyl gravity and the torsional teleparallel theory of Maluf and Faria \cite{Maluf:2011kf}. Both our and the torsional approach have a clear advantage with respect to the Weyl theory due to the absence of higher derivatives and hence naturally avoid Gauss-Ostrogradsky ghosts. Due to this property, it is obvious that the theory given by the Lagrangian \eqref{eq:Geom Conf Lagr} is not equivalent to Weyl gravity, since it is generally not possible to establish an equivalence between theories with second-order field equations on one side and fourth-order field equations on the other side. 

We can also compare our theory directly with the torsional approach of Maluf and Faria \cite{Maluf:2011kf}. The Lagrangian \eqref{eq:Geom Conf Lagr} resembles the torsional Lagrangian of torsional conformal gravity, i.e. it is quadratic in the non-metricity scalar and its form is fixed by conformal invariance. In general, even by counting the free parameters of the two theories, it can be seen that they are distinct. In the torsional approach we have three parity-preserving quadratic invariants and hence three free parameters in the torsion scalar. The conformal invariance gives us one constraint which leaves us with two free parameters \cite{Maluf:2011kf}. On the other hand, in the symmetric teleparallel framework we have five parity-preserving quadratic invariants and the conformal invariance introduces two independent constraints on parameters and we are left with three free parameters \eqref{eq:A_B_C}. It is generally not possible to establish a one-to-one mapping between theories with a different number of 
free parameters.

We can also highlight that Symmetric Teleparallel Gravity seems to be a more natural framework, since the conformal transformations naturally act directly on the metric. This is in contrast with the torsional Teleparallel Gravity, where the fundamental 
variable is the tetrad, on which the conformal transformation acts. This transformation is then related to the metric through the tetrad rather than the direct approach presented here.

Therefore, our symmetric teleparallel approach combines attractive features of both the Weyl gravity and the torsional teleparallel conformal gravity \cite{Maluf:2011kf}, since it has the metric as the fundamental variable and also avoids the problem of higher order field equations. 

Let us conclude with the suggestion that it would be interesting to explore the relation between all three distinct approaches to conformal gravity and understand the possible links among them. General Relativity can be cast into three different formulations using three different geometrical frameworks (GR, TEGR, STEGR); in fact this is the so-called Trinity of Gravity \cite{BeltranJimenez:2019tjy}. On the other hand, it seems that it is not possible to establish the same equivalency in the case of conformal gravity, as each geometrical framework seems to provide a distinct theory.

\section*{Acknowledgements}
The authors would like to acknowledge networking support by the COST (European Cooperation in Science and Technology) Action GWverse CA16104. This article is based upon work from CANTATA COST  action CA15117, EU Framework Programme Horizon 2020.

%%%%%%%%%%%%%%%%%%%%%%%%%%%%%%%%%%%%%%%%%%%% Appendixes

\appendix
%dummy comment inserted by tex2lyx to ensure that this paragraph is not empty

\section{Inverse conformal transformations for non-metricity invariants\label{Conf_irr_app}}

The conformal transformation introduced at the level of the metric
in (\ref{Conf_trans_def}) results in the irreducible conformal
transformations presented in Eqs.~(\ref{Conf_irr_1})-(\ref{Conf_irr_5}).
The separate non-metricity tensors have the following inverse conformal
transformation relations:
\begin{align}
Q_{a\mu\nu} & 
=\Omega^{-2}\left(\widetilde{Q}_{\alpha\mu\nu}-2\widetilde{g}_{\mu\nu}\widetilde{\partial}
_{\alpha}\ln\Omega\right)\\
Q^{\alpha\mu\nu} & 
=\Omega^{4}\left(\widetilde{Q}^{\alpha\mu\nu}-2\widetilde{g}^{\mu\nu}\widetilde{\partial}^
{\alpha}\ln\Omega\right)\\
Q_{\mu} & =\widetilde{Q}_{\mu}-8\widetilde{\partial}_{\mu}\ln\Omega\\
\bar{Q}_{\mu} & =\widetilde{\overline{Q}}_{\mu}-2\widetilde{\partial}_{\mu}\ln\Omega.
\end{align}
These inverse transformations can be finally combined in order to produce the inverse conformal transformations of the irreducible components of STG, which are 
\begin{align}
Q_{\alpha\mu\nu}Q^{\alpha\mu\nu} & 
=\Omega^{2}\Big(\widetilde{Q}_{\alpha\mu\nu}\widetilde{Q}^{\alpha\mu\nu}-4\widetilde{Q}^{
\alpha}\widetilde{\partial}_{a}\ln\Omega \nonumber\\
& \ \ \ \ \ \ \ \ \,  + 16\widetilde{\partial}^{\alpha}
\ln\Omega\widetilde{\partial}_{\alpha}\ln\Omega\Big),\\
Q_{\alpha\mu\nu}Q^{\mu\alpha\nu} & 
=\Omega^{2}\Big(\widetilde{Q}_{\alpha\mu\nu}\widetilde{Q}^{\mu\alpha\nu}-4\widetilde{\bar
{Q}}^{\alpha}\widetilde{\partial}_{a}\ln\Omega \nonumber\\
& \ \ \ \ \ \ \ \ \,  + 4\widetilde{\partial}^{\alpha}\ln\Omega\widetilde{\partial}_{\alpha}\ln\Omega\Big),\\
Q_{\alpha}Q^{\alpha} & 
=\Omega^{2}\Big(\widetilde{Q}_{\alpha}\widetilde{Q}^{\alpha}-16\widetilde{Q}^{\alpha}
\widetilde{\partial}_{a}\ln\Omega \nonumber\\
& \ \ \ \ \ \ \ \ \,  + 64\widetilde{\partial}^{\alpha}\ln\Omega\widetilde{\partial}_{\alpha}\ln\Omega\Big),\\
\bar{Q}_{\alpha}\bar{Q}^{\alpha} & 
=\Omega^{2}\Big(\widetilde{\bar{Q}}_{\alpha}\widetilde{\bar{Q}}^{\alpha}-4\widetilde{\bar
{Q}}^{\alpha}\widetilde{\partial}_{a}\ln\Omega \nonumber\\
& \ \ \ \ \ \ \ \ \,  + 4\widetilde{\partial}^{\alpha} \ln\Omega\widetilde{\partial}_{\alpha}\ln\Omega\Big),\\
\bar{Q}_{\alpha}Q^{\alpha} & 
=\Omega^{2}\Big(\widetilde{\bar{Q}}_{\alpha}\widetilde{Q}^{\alpha}-8\widetilde{\bar{Q}}^{
\alpha}\widetilde{\partial}_{a}\ln\Omega-2\widetilde{Q}^{\alpha}\widetilde{\partial}_{a}
\ln\Omega \nonumber\\
& \ \ \ \ \ \ \ \ \,  + 16\widetilde{\partial}^{\alpha}\ln\Omega\widetilde{\partial}_{\alpha}\ln\Omega\Big).
\end{align}

\end{document}